# Improving specificity and axial spatial resolution of refractive index imaging by exploiting uncorrelated subcellular dynamics

Hervé Hugonnet[1,2], HyunJun Han[1,2], Weisun Park[1,2], and YongKeun Park[1,2,3*]

[1] *Department of Physics, Korea Advanced Institute of Science and Technology (KAIST), Daejeon 34141, South Korea;*

[2] *KAIST Institute for Health Science and Technology, KAIST, Daejeon 34141, South Korea;*

[3] *Tomocube Inc., Daejeon 34109, South Korea*

*corresponding authors: Y.K.P (yk.park@kaist.ac.kr)

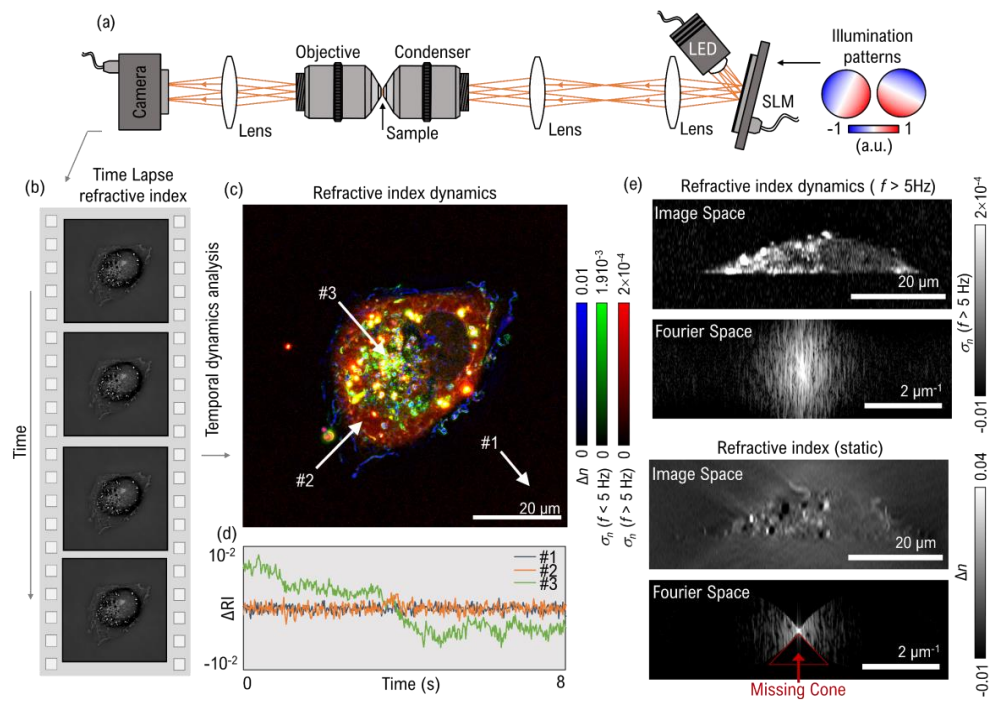

**ABSTRACT**

Holotomography, a three-dimensional quantitative phase imaging technique, presents an innovative, non-invasive approach to studying biological samples by exploiting the refractive index as an intrinsic imaging contrast. Despite offering label-free quantitative imaging capabilities, its potential in cell biology research has been stifled due to limitations in molecular specificity and axial resolution. Here, we propose and experimentally validate a solution to overcome these constraints by capitalizing on the intrinsic dynamic movements of subcellular organelles and biological molecules within living cells. Our findings elucidate that leveraging such sample motions enhances axial resolution. Furthermore, we demonstrate that the extraction of uncorrelated dynamic signals from refractive index distributions unveils a trove of previously unexplored dynamic imaging data. This enriched dataset paves the way for fresh insights into cellular morphologic dynamics and the metabolic shifts occurring in response to external stimuli. This promising development could broaden the utility of holotomography in cell biology.

# INTRODUCTION

Quantitative phase imaging (QPI) leverages the refractive index (RI) of samples as an intrinsic imaging contrast[1]. Among QPI techniques, holotomography (HT), also called optical diffraction tomography, stands out for its capacity to depict cellular and subcellular morphology in high-resolution 3D[2-4]. HT's inverse scattering approach enables rapid and accurate reconstructions of the 3D RI distribution of individual cells. Its label-free and quantitative imaging prowess has quickly propelled its application across diverse biological domains, from cell biology and microbiology[5-9], to the niches of phase separation in biology[10-12], lipid imaging[13-15], immunology[16], preclinical research[17], and regenerative medicine[18].

Nonetheless, HT encounters challenges due to its limited specificity and subpar axial resolution[3,19]. Although RI has a linear relationship with the concentration of biomolecules[20,21], it does not generally offer molecular-specific information. An exception to this is lipid droplets[22,23], which exhibit distinctly high RI values in contrast to cytoplasm. Moreover, HT is hampered by the limited numerical apertures (NAs) of both illumination and detection optics, preventing the modulation and collection of a certain fraction of light information. Consequently, this restricts access to specific spatial frequency information, a limitation commonly referred to as the 'missing cone problem', which greatly limits the sectioning ability of HT[19]. HT differs from modern fluorescence imaging methods where light emitted from varied positions within the sample is incoherent and uncorrelated, thereby enabling the use of super-resolution techniques[24,25], in HT the light scattered by the RI is coherent to the illumination, causing interference between light scattered at different positions inside the sample, resulting into the missing cone problem.

To address these challenges, our study delves into the potential of leveraging uncorrelated subcellular motion probed by measuring the dynamics of 3D RI distributions of cells to enhance both resolution and specificity in HT. Sample dynamics have already been widely exploited in reflection microscopy techniques for the measurement of blood flow[26,27], reduction of speckle noise and increasing the imaging content[28-33] or even measurement of ultrasound wave[34]. In particular when looking at high-resolution imaging methods, several algorithms have been developed for optical coherence tomography (OCT)[32,35], full-field OCT[29,36,37] or interferometric scattering[38]. Application to transmission microscopy remains however limited[33,39-41], especially the imaging of 3D dynamics has been restricted[33] due to the high imaging speed requirement, which is incompatible with multi-image acquisition schemes used for volumetric imaging, especially in HT with angular scanning of the illumination.

Here, we leverage a recent advancement in low-coherence HT, referred to as Single Section RI Deconvolution (SISRID)[42]. This method facilitates high-speed imaging of the RI at the current focal plane by eliminating the need for full 3D volume acquisition. By measuring the RI dynamics at a specific focal plane and subsequently employing axial scanning, we effectively capture the three-dimensional high-speed dynamics of the sample. Data is then analysed by employing spatiotemporal filtering of the RI data. This approach enables us to maximize the image content and resolution. As a capstone to our research, we demonstrate potential biomedical applications of our method by highlighting alterations in sample dynamics in response to drug administration. These findings not only validate our approach but also illuminate new pathways for the application of HT in biomedical studies.

## Results

**Improvement of axial resolution**

In order to illustrate the extraction of dynamic signals from time-lapse 3D RI tomograms, we employed the SISRID method on live, unlabeled biological cells (See Methods for details, Fig. 1(a)). We obtained 2D RI tomograms sections of a COS7 cell every 20 millisecond for a duration of 8 seconds (See Fig. 1(b)) repeating the operation at every z position with a scanning step of 0.5 μm over 28 μm to obtain a 3D tomogram.

From the acquired time-lapse 3D RI tomogram, we extracted three 3D dynamics maps by conducting analysis utilizing the inherent dynamic movements of subcellular organelles and biological molecules within living cells (See Methods). Data can be visualised using three quantity: RI contrast, $\Delta n$, and maps for slow and fast dynamics. In this study, we defined the fast dynamics as '> 5 Hz', which corresponded to data standard deviation high-pass filtered at 5 Hz and 2.3 $\mu m^{-1}$. On the other hand, the slow dynamics, labelled as '<5 Hz', corresponded to data standard deviation low-pass filtered at 5 Hz. These three 3D maps were then combined into a single coloured 3D image for better visualization. We assigned the RI contrast to the blue channel; the dynamic signal derived from slow ($f < 5$ Hz) dynamics to the green channel; and the dynamic signal derived from fast ($f > 5$ Hz) dynamics to the red channel.

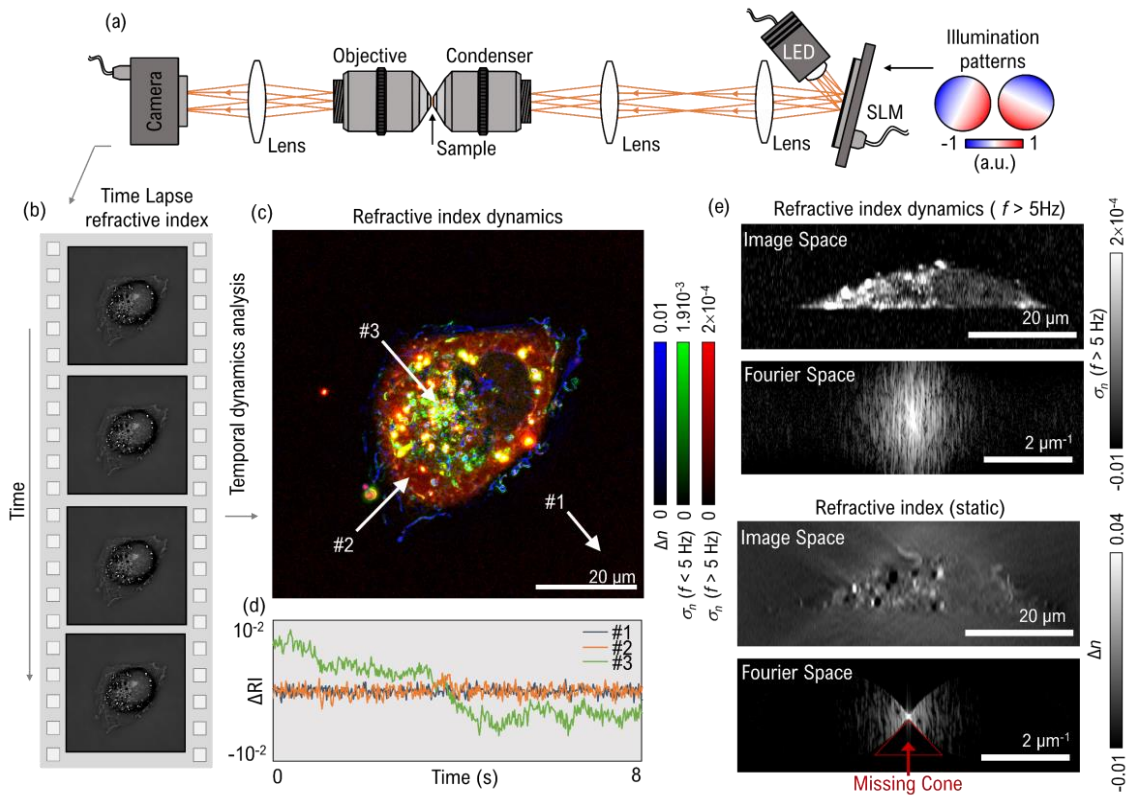

**Figure 1 | Reconstruction result and resolution improvement, (a)** Imaging system used for measuring time-lapse RI acquisitions **(b)** High-speed time-lapse RI image of a live unlabelled COS7 cell. **(c)** RI dynamics image composed of three color-coded channel information: RI contrast (the blue channel); dynamic signal extracted from slow ($f < 5$ Hz) dynamic RI variances; and dynamic signal extracted from fast ($f > 5$ Hz) dynamic RI variances. **(d)** Variation of the refractive index as a function of time at different point in the sample. (e) Axial sections of the dynamic and static

The composite RI dynamic image is shown in Fig. 1(c). The time-dependent fluctuation of RI can also be visualised. Three representative positions were selected and their RI contrast traces are shown in Fig. 1(d).

Imaging of uncorrelated dynamics improves optical sectioning (See Methods for derivation). The improvement in axial resolution is experimentally demonstrated by comparing the tomograms of the same COS7 cell visualized with a 3D static RI tomogram and with the 3D dynamic analysis tomogram. As displayed in Fig. 1(d), the 3D RI static tomogram is axially elongated along the optical axis. Additionally, its Fourier transform reveals the missing cone area, indicating the absence of crucial axial information. In contrast, the RI dynamics tomogram clearly depicts the cell boundary, the improvement in axial resolution is especially noticeable at the interface between the cell and the substrate. Crucially, its Fourier representation demonstrates a substantial extension of the spatial bandwidth in the axial direction.

**Three-dimensional dynamic imaging with improved specificity**

To further explore dynamic RI tomography, we analyse in Figure 2 each component of 3D dynamic image of COS7 cells. Low-frequency and high-frequency dynamic RI tomograms as well as the static RI tomogram, and the composite RGB image are displayed in Figs. 2(a)-2(d). Notably, the RI channel (Fig. 2(c)) was imaged separately using 3D deconvolution[43], distinct from the 2D deconvolution[42] employed for the dynamic and composite image. This approach was chosen to mitigate imaging artefacts associated with out-of-focus SISRID images (Further details can be found in the supplementary materials).

The composite image highlights important features of the live cells. The cell cytoplasm is rendered in red (Fig. 2d) due to a robust signal in the fast-dynamic channel. The overall cell shape is also well illustrated, including the cell and nucleus membranes. The cell membrane's edge is visualized via the static RI contrast $\Delta n$ in blue and showcases behaviours related to membrane ruffling. In most areas of the cytoplasm, subcellular organelles display slow dynamics, which are indicated by the green colour channel. Slower dynamics is expected due to the larger size of organelles when compared with free floating protein in the cytoplasm. The nucleus membrane and nucleoplasm can be discerned from the static RI contrast. Nucleoplasm showcases fast yet weak dynamics, evident from its comparison with the static background. Nucleoli show a high refractive index but dynamics even weaker than the nucleoplasm. These finding are in concordance with another study[39]. The cytoplasm predominantly exhibits dynamic signals with an uneven distribution of dynamics, which suggests the movement of subcellular organelles and vesicles. Among them, lipid droplets are noticeably visible in both fast and slow dynamic images, leading to their yellow appearance in the composite image. The area of the cytoplasm located to the left of the nucleus, highlighted by the white arrow (Fig. 1c. #3), displays slow dynamics, setting it apart from other parts of the cytoplasm. This could suggest that these areas are representative of the endoplasmic reticulum.

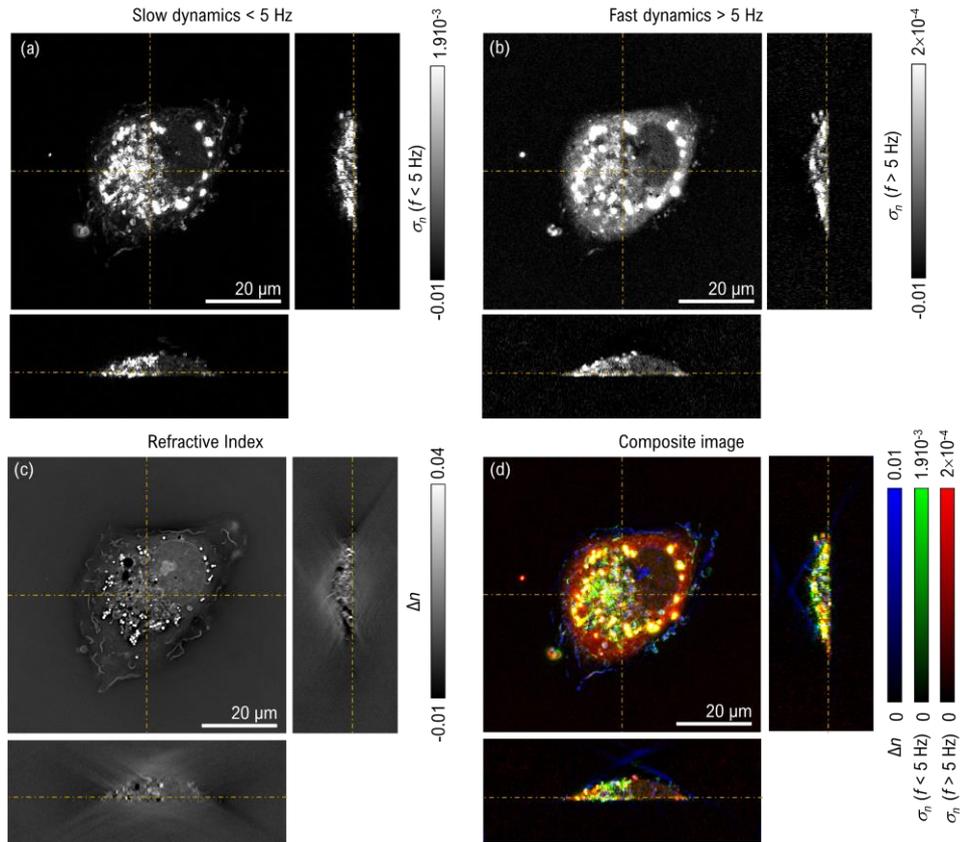

**Figure 2 | 3D dynamic image. (a)** Slow dynamics at a frequency lower than 5 Hz **(b)** Fast dynamics at a frequency higher than 5 Hz (c) refractive index image **(c)** RGB image with each channel corresponding to one modality.

In terms of axial resolution, as anticipated in the methods section, the modified PSF enables both slow and fast dynamic images to precisely capture the cell morphology (Figs. 2a and 2b). Yet the refractive index image faces difficulties in clearly portraying the distinct boundary between the cell and the medium in the vertical direction, owing to the missing cone issue (Figs. 2c, 1e)[19,34]. Finally, the importance of spatial filtering is exemplified in Fig. S2. and derived in the methods section. Without spatial filtering, colour contrast is notably diminished.

**Comparison with fluorescence imaging**

To deepen our understanding of the biological implications of the retrieved dynamic RI imaging, we carried out both dynamic 3D RI imaging and 3D fluorescence imaging of the same HeLa cells at high density (see Methods). The findings are presented in Fig. 3.

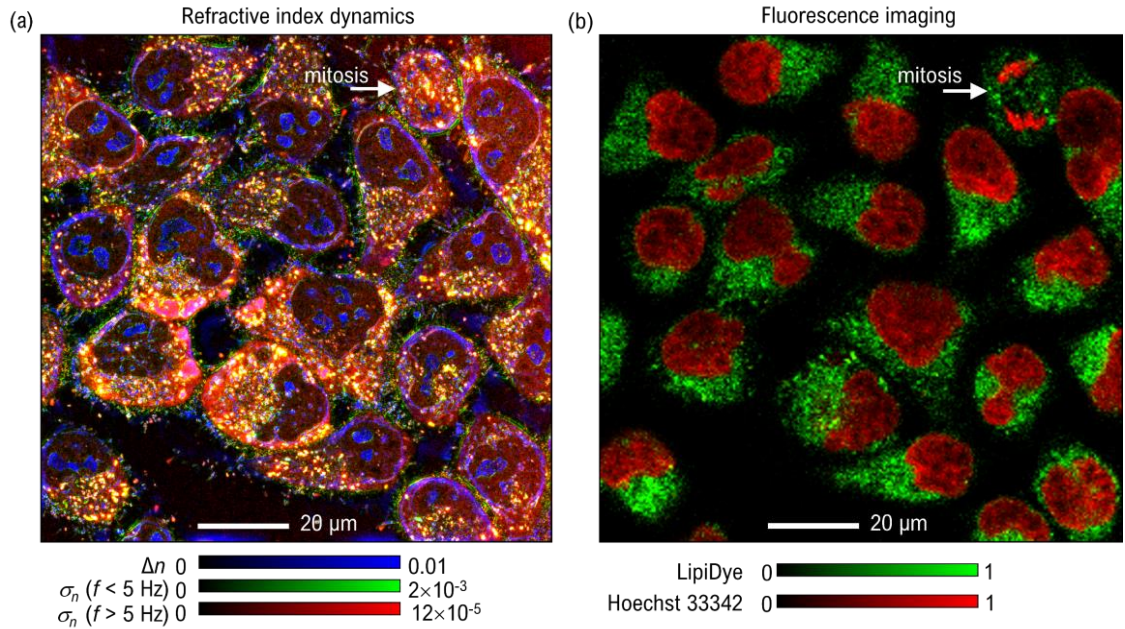

**Figure 3 | Comparison of the same cells using (a) dynamics imaging and (b) fluorescence imaging.** A cell undergoing mitosis is present in the top right corner.

Overall, the nuclei correspond well with the darker parts of the cells in the dynamic image. An interesting exception is a cell in the top right corner of the image, which is currently undergoing mitosis. In this cell, the nucleus exhibits a higher level of dynamics compared to the cytoplasm, likely due to the ongoing nuclear division. The distribution of bright yellow spots in the dynamic image aligns with the lipid droplet fluorescent marker. In most cells, lipid droplets are located in the cytoplasm but are absent from the nucleus. However, in the mitotic cell, lipid droplets can be observed in the nucleus in both the dynamic and fluorescence images. Although we utilized different microscopes for each modality in this study, a multimodal microscope could be developed in the future to leverage both techniques.

**Change in subcellular dynamics in response to drug delivery**

To demonstrate utility of the signal we further experimented by time-lapsed imaging of COS7 cells before and after exposure to 50 μM of tamoxifen, a cancer drug, for 3 hours (Fig. 4). This time, we limited imaging to 2D acquisition since this enables a total acquisition time of 8 seconds, enabling imaging of multiple samples. Images show significantly altered cell metabolism after tamoxifen administration. For example, before drug admission the dynamic of the cell nucleoli was significantly lower than the cytoplasm, while after drug admission both show a similar dynamic strength. Additionally, to an overall change in the shape of the cell, the number of lipid droplet was also greatly reduced, which can be directly noticed in the composite image by the absence of yellow.

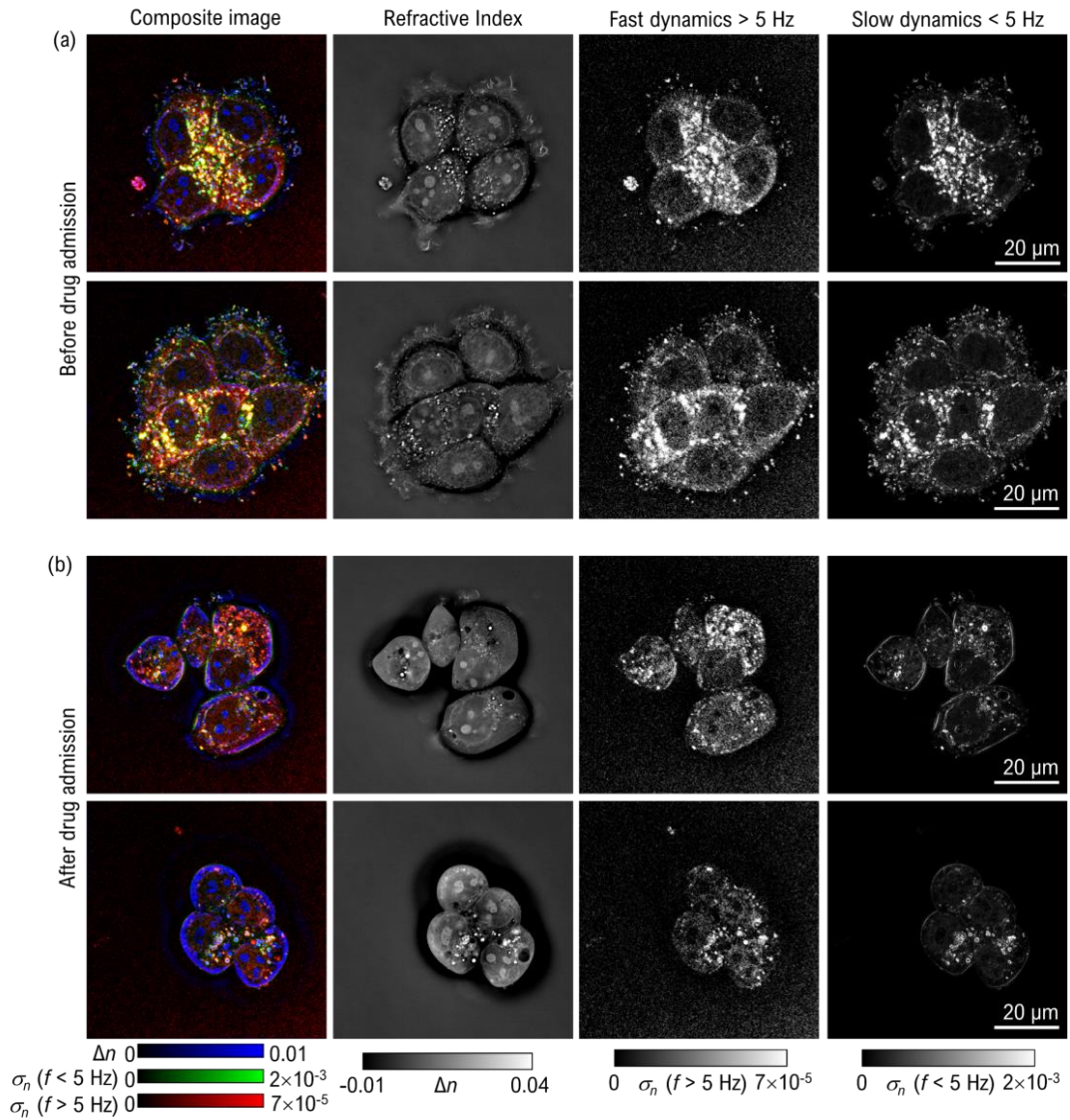

**Figure 4 | Change in sample dynamics as a result of drug admission.** **(a)** Cells before drug admission **(b)** Cells after drugs admission. Note that **(a)** and **(b)** do not represent the same cells.

## Discussion

In this research, we have proposed and validated a methodology for the visualization of 3D refractive index dynamics, offering a way to examine and decipher the intrinsic dynamics within living cells. This approach successfully provides high-resolution, three-dimensional imaging of cellular dynamics, by integrating low-coherence HT, spatiotemporal filtering, and image processing techniques.

Our methodology addresses the enduring missing cone problem and decreases associated imaging artefacts by retrieval of dynamic signals from time-lapse 3D RI tomograms. This method reshapes the point spread function reducing missing cone artefacts appearing in the visualization of static RI tomogram and leading to a significant improvement in axial resolution. This key advancement enables us to visualize and understand subcellular structures and dynamics along the axial direction with minimal requirement for external labels.

The various experiments conducted on biological samples like COS7 cells, HeLa cells, and cells treated with tamoxifen, a cancer drug, highlight the considerable potential of our method in biological and biomedical research[44]. The ability to concurrently visualize slow and fast dynamics within living cells delivers invaluable insights into cellular function and their responses to external factors, such as drug interventions.

Although the present method heralds numerous applications in cell biology, medical research, and biotechnology, particularly in enhancing the specificity of long-term label-free imaging, we acknowledge certain limitations. The acquisition speed of our current setup is limited, restricting the temporal resolution of the 3D dynamic imaging. However, future improvements with faster or more sensitive image sensors and the use of multicore processing units for rapid image reconstruction may significantly enhance the imaging speed, facilitating real-time monitoring of cellular processes.

In terms of molecular specificity, our technique is limited compared to fluorescence-based imaging. However, the exploitation of spatiotemporal dynamic analyses in our method aided with machine learning approaches may offer a level of specificity by identifying distinctive dynamic patterns associated with various molecular activities[45-47].

We also envision incorporating dynamic RI tomography with other imaging techniques, like 3D fluorescence or Raman imaging, in a multimodal microscope setup[48-50]. This would increase the depth and variety of information available to researchers, supporting a more comprehensive study of biological processes.

An exciting prospect of our approach is the potential to develop novel imaging analysis parameters based on RI dynamics. These parameters may offer unique insights into cellular function and behaviour, including responses to drug treatments, cell cycle events such as cell division, disease progression, and other cellular events[51-53].

In this work, we have assumed the thermally driven equilibrium dynamics to model the dynamics of cells measured via HT. However, there are non-equilibrium dynamics and also active motions driven by motor proteins, further advancements in the interpretation of dynamic RI tomogram would lead to more in-depth understanding of subcellular dynamics[54].

In conclusion, our research has introduced an effective approach for studying cellular dynamics, providing a fresh perspective into the inner workings of living cells. We anticipate that dynamic RI tomography will serve as a powerful tool for biologists and medical researchers, ushering in new discoveries and deeper understanding of cellular processes and disease pathogenesis.

## Materials and Methods

**High-speed imaging of the refractive index.** To exploit dynamic RI signals, enhance specificity and axial spatial resolution, swift acquisition of time-lapse RI tomograms is important. Volumetric RI imaging in HT primarily hinges on two acquisition schemes. The first, called optical diffraction tomography, modulates the illumination angles while recording diffracted field images[2], and the second involves recording diffracted intensity while refocusing the sample[55]. Both methods necessitate acquiring multiple 2D image data, typically ranging from 50 to a few hundred images, to reconstruct a 3D RI tomogram. Therefore, blurring fast sample motion.

The utilization of subcellular motion in live biological cells to boost contrast and specificity is a well-established approach in OCT[28-32]. However, subcellular dynamics, are often measured on the order of 10 Hz, which remains predominantly incompatible with most HT methods[36]. This discrepancy poses a challenge that our research aims to address.

In this work, we utilized the SISRID method[42] for high-speed RI tomograms. The SISRID makes use of engineered illumination to enable reconstruction of the in-focus RI at high speed. The SISRID requires only four images to recover the RI slice image at a specific focal plane, and opens the possibility for imaging fast subcellular motion.

SISRID operates on the principles of partially-coherent optical diffraction tomography. It computes the RI by deconvolving the transmitted intensity through a sample. The sample is illuminated by a uniform, spatially incoherent beam whose intensity can be modulated in the pupil plane. Image formation can be described as the convolution of the real and imaginary parts of the scattering potential index with two point spread functions (PSFs)[42]:

$$(I_{in} - I_{out})/I_{out} = H_A * V_{img} + H_P * V_{real}, \tag{0.1}$$

where $I_{in}$ and $I_{out}$ are the recorded intensity in presence and absence of sample, $V_{img}$, $V_{real}$ the real and imaginary part of the scattering potential and $H_A$, $H_P$ are the PSFs defined as

$$\begin{aligned} H_P &= -128\pi^4 \cdot \mathrm{Im}\left(G^* \cdot \mathcal{F}[\mathcal{F}[G] \cdot k_z \cdot \rho]\right) \\ H_A &= -128\pi^4 \cdot \mathrm{Re}\left(G^* \cdot \mathcal{F}[\mathcal{F}[G] \cdot k_z \cdot \rho]\right) \end{aligned}, \tag{0.2}$$

where G is the green function of the optical system, $\rho$ the intensity distribution at the pupil plane, $\mathcal{F}$ the 3D Fourier transform and $k_z$ the vertical component of the wavevector. In SISRID, the illumination $\rho$ is optimized so that $H_A$ is constant in the vertical direction. Owing to this property of the transfer function, deconvolution along the vertical axis is not needed, and the refractive index can be directly obtained from two-dimensional data. This feature, combined with our implementation, allowed us to achieve an imaging speed of 50 Hz for single section imaging.

**Optical setup.** To apply the SISRID method, a custom-built optical imaging system was used (Fig. 1(a)). A light-emitting diode (LED) with a centre wavelength of 695 nm was used as the low-coherence illumination source. The distribution of illumination intensity was managed in the Fourier plane using an intensity spatial light modulator (SLM; Forth Dimension Display M150). This was then projected onto a sample through a water-immersion condenser lens (UPLSAPO60XW, NA 1.2, Olympus). Diffracted light from the sample was gathered via a water-immersion objective lens (UPLSAPO60XW, NA 1.2, Olympus), projected onto the image plane, and then captured using a CMOS camera (ORX-10G-71S7M-C, 250 fps at 2×2 binning, FLIR).

**Motion dynamics depending on particle size.** Cellular organelles perpetually exhibit motion, governed by both thermal Brownian motion and active non-equilibrium movements. In our study, we postulate that the dominant dynamics, as captured by measuring the dynamic 3D RI signals, are principally thermally-driven random Brownian motions. This physical phenomenon, which exerts a more profound impact on smaller particles compared to larger ones, forms the foundation of numerous microscopy and measurement techniques. These techniques are specifically designed to study particles whose sizes surpass the resolution limit. One of the most prevalent methods is dynamic light scattering[56-58], which employs autocorrelation of intensity to determine particle size and viscosity within a colloidal solution. Our technique further reinforces the potential and versatility of leveraging particle dynamics within microscopy studies.

By taking a time-lapse image $I_a(\vec{r},t)$ of a small particle of diameter a, we can analyse its spatial and temporal behaviour through the use of the Fourier transform. One useful property of the Fourier transform is the conservation of total signal intensity across both the Fourier and image space. As a result, Fourier analysis enables us to predict how different filters will influence the signal intensity emanating from particles of varying sizes. We can further make use of dynamic light scattering theory[56] since autocorrelation of the intensity is linked to the power spectrum by the Fourier transform[59]. For example, for a spherical particle of diameter $a$, the following relationship can be derived (see the supplemental material for the detail):

$$\left|\mathcal{F}_x[\mathcal{F}_t[I_a(\vec{r},t)](\omega)](\vec{k})\right|^2 = P_a(\vec{k},\omega)$$
$$\simeq 9\left|\frac{\sin(\|\vec{k}\|_2 a) - \|\vec{k}\|_2 a \cos(\|\vec{k}\|_2 a)}{(\|\vec{k}\|_2 a)^3}\right|^2 \frac{2a/\|\vec{k}\|_2^2 (k_B T/3\pi\eta)}{(\omega a/\|\vec{k}\|_2^2 (k_B T/3\pi\eta))^2 + 1}, \quad (0.3)$$

where $\vec{k},\omega$ are the spatial and temporal frequency coordinate and $k_B, T, \eta$ are respectively the Boltzmann constant, temperature, and viscosity. This formula shows that smaller particles have both a broader spatial and temporal spectrum. So that high-pass filtering the spatial and temporal dimensions will select the signal coming from smaller particles. The filtering efficiency for high-pass temporal filtering can be derived (c.f., Supplementary) as $\frac{P_{filtered}(a_1)}{P_{filtered}(a_2)} \propto \left(\frac{a_2}{a_1}\right)^1$ where $P_{filtered}(a)$ is the filtered signal strength produced by a particle of size a. While for spatial filtering $\frac{P_{filtered}(a_1)}{P_{filtered}(a_2)} \propto \left(\frac{a_2}{a_1}\right)^4$ showing that spatial filtering is more efficient at selection smaller particles than temporal filtering. This demonstrates that in contrast to reflection technique such as OCT where signal already corresponds to high spatial frequency, in HT spatial filtering is essential in order to select small particles, since due to the transmission geometry low frequency information about the refractive index is also retrieved.

While this theoretical formulation provides insights into the underlying mechanisms driving Brownian motion within cells, the practical application is more complex. Predominantly, most organelles within a cell do not adopt a spherical shape; rather, they tend to have more complex, often elongated or textured forms, rendering spatial filtering often less effective compared to theoretical expectations. Additionally, viscosity within cells has been measured utilizing fluorescence correlation spectroscopy [60]. These experiments indicate that cell viscosity is profoundly influenced by particle size, increasing proportionally with it. This finding suggests that temporal filtering might offer more effective particle size filtering than predicted in our theoretical framework.

These intricate characteristics render it challenging to determine an optimal filter based on theoretical derivations alone. Nonetheless, dynamic information can still provide valuable qualitative insights about a sample[61,62]. In practical terms, filtering characteristics are determined by striking a balance between the signal-to-noise ratio and the selectivity of the filter. For instance, in dynamic full-field OCT, a high-pass temporal filter around 5 Hz is commonly used[35], which we found suitable for our application as well. As for the spatial filter, a larger high-pass filter leads to increased specificity but at the expense of a reduced signal-to-noise ratio. Consequently, we opted for a filter of 2.3 μm$^{-1}$, as it offered an effective balance between image content and signal-to-noise ratio.

**Total birefringence regularization for the multiple scattering model.** In the last subsection, we demonstrated how spatiotemporal filtering can be used to filter particles based on their size. But we still need to analyse the filtered time-lapse signal. The raw filtered signal looks like speckle noise due to the interference of the signals coming from other particles. Although various analysis methods are available[36,37], a common method to analyse this signal is to use the temporal variance [29,39,41]. If the signal $S(x,t)$ is expressed as the sum of signals $S_i(x,t)$ coming from the different subcellular particles $S(\vec{r},t) = \sum_i S_i(\vec{r},t)$, then due to the random motion of the particles the variance of the total signal can be expressed as the sum of each particle's variance (derivation provided in Supplementary),

$$Var\left(S(\vec{r},t)\right) = \sum_i Var\left(S_i(\vec{r},t)\right). \quad (0.4)$$

Suppose the particle i has a scattering potential distribution $V_i(\vec{r})$ and the high-pass filter is denoted as $F_{High}$ then we can reformulate this expression as

$$Var\left(S(\vec{r},t)\right) = \sum_i p_i(\vec{r}) \otimes \left[V_i(\vec{r}) \otimes PSF_{High}(\vec{r})\right]^2 P_{temporal}(a), \quad (0.5)$$

where $p_i(\vec{r})$ is the occupancy probability of the particle $i$ or particle density, and $PSF_{High}(\vec{r}) = PSF(\vec{r}) \otimes F_{High}$ is the PSF after spatial high-pass filtering. The image can finally be understood as particle density imaged with a PSF $\left[V_i(\vec{r}) \otimes PSF_{High}(\vec{r})\right]^2$ dependent on the particle shape $V_i(\vec{r})$ and the microscope resolution. For particle larger than the diffraction limit the resolution is then approximately equal to the particle size while for sub diffraction size particle the resolution is diffraction limited. This also hints that sample composed of large particle might have some speckle noise due to the non-uniform PSF depending on the particle shape, while spatiotemporally filtered data which mainly retain information about small particles should be less noisy and higher resolution.

An essential characteristic of this analysis is the squaring of the PSF, which allows access to new spatial frequencies compared to the original PSF. One significant challenge in 3D transmission microscopy is axial sectioning. It is difficult to determine the sample's position along the propagation direction of the illumination, as image formation in optical diffraction tomography can be perceived as a projection [19,50]. This leads to artefacts and image deformation when examining cross-sections, the missing cone problem. The squaring of the PSF significantly mitigates the missing cone problem and reduces associated artefact's (Fig. 1e).

**Sample preparation.** Cells were maintained in Dulbecco's modified Eagle's Medium (DMEM, Gibco, CA, USA; ATCC, 30-2002) supplemented with 10% fetal bovine serum (Thermo Fisher Scientific Inc.), 1% (v/v) penicillin/streptomycin (Thermo Fisher Scientific Inc.) at 37°C in a 5% CO2 incubator. All samples were loaded into imaging dishes (TomoDish, Tomocube Inc.) at a density of $1\times10^{-5}$ cells/ml.

Hela cells (CCL-2, ATCC) were cultured under the same conditions as COS7 cells (CRL-1651, ATCC). Since dynamics show the nucleus and lipid droplets with high contrast we stained Hela cells with Hoechst 33342 (H3570, Invitrogen) and LipiDye Lipid Droplet Green (Funakoshi, FDV-0010). Cell dynamics where first imaged, then the cells were rapidly fixed using 2% PFA and 0.05 GA in PHEM buffer and were then transferred to a confocal microscope (A1R-HD25, Nikon) for 3D fluorescence imaging.

**Acknowledgements**

This work was supported by National Research Foundation of Korea (2015R1A3A2066550, 2022M3H4A1A02074314), Institute of Information & communications Technology Planning & Evaluation (IITP; 2021-0-00745) grant funded by the Korea government (MSIT), KAIST Institute of Technology Value Creation, Industry Liaison Center (G-CORE Project) grant funded by MSIT (N11230131).

**Author contributions**

H. Hugonnet performed and analysed the experiments, developed a theory and wrote the manuscript. H. Han performed and analysed the experiments. W.P provided the samples and wrote the manuscript. Y.P. supervised the project and wrote the manuscript.

**Conflict of interest**

H. Hunonnet and Y.K. Park have financial interests in Tomocube Inc., a company that commercializes holotomography instruments and is one of the sponsors of the work.

**Data availability**

Full-resolution images presented in this study are available on request. Correspondence and requests for materials should be addressed to Y.P.